\documentclass[preprint]{aastex62}

\usepackage{amsmath}

\usepackage{graphicx}
\received{}
\revised{}
\accepted{}

\shorttitle{A Statistical Study of Solar Radio Type III Bursts and Space Weather Implication}
\shortauthors{Ndacyayisenga et al.}

\begin{document}

\title{A Statistical Study of Solar Radio Type III Bursts and Space Weather Implication}

\correspondingauthor{Theogene Ndacyayisenga}
\email{ndacyatheogene@gmail.com}
\author{Theogene Ndacyayisenga}
\affiliation{University of Rwanda - College of Science and Technology, Kigali, P.O.BOX 3900, Rwanda}
\affiliation{Department of Physics, Mbarara University of Science and Technology, P. O. Box 1410, Mbarara, Uganda.}
\author{Jean Uwamahoro}
\affiliation{University of Rwanda, College of Education, P.O. BOX 55, Rwamagana -- Rwanda.}

\author[0000-0002-1192-1804]{K. Sasikumar Raja}
\affil{Indian Institute of Astrophysics, II Block, Koramangala, Bengaluru - 560 034, India.}

\author{Christian Monstein}
\affiliation{Istituto Ricerche Solari Locarno (IRSOL), Via Patocchi 57, 6605 Locarno Monti, Switzerland.}




\begin{abstract}
Solar radio bursts (SRBs) are the signatures of various phenomenon that happen in the solar corona and interplanetary medium (IPM). 
In this article, we have studied occurrence of Type III bursts and their association with the Sunspot number. This study confirms that 
occurrence of Type III bursts correlate well with Sunspot number. Further, using the data obtained using e-CALLISTO network, 
we have investigated drift rates of isolated Type III bursts and duration of the group of Type III bursts. 
Since Type II, Type III and Type IV bursts are signatures of solar flares and/or CMEs, we can use the radio observations to predict space weather hazards. In this article, we have discussed two events that have caused near Earth radio blackouts. 
Since e-CALLISTO comprises more than 152 stations at different longitudes, we can use it to monitor the radio emissions from the solar 
corona 24 hours a day. Such observations play a crucial role in monitoring and predicting space weather hazards within few minutes to hours of time.\\

\end{abstract}



\section{Introduction} \label{sec:intro}
\label{sec1}
\noindent Severe disturbances in the near Earth's  space environment are caused by solar activity. The eruptive events like solar flares and coronal 
mass ejections (CMEs) are often associated with radio emissions which originate in the solar corona. Such radio emissions or solar radio bursts (SRBs) describe various physical processes that happens in the solar corona and interplanetary medium (IPM). Based on the drifting 
rates and morphology as seen in the dynamic spectrograms, SRBs are primarily classified into five classes vis. Type - I to V \citep{Wild1950, Wild1963}. 
In this article, we study fast drifting Type III bursts, slow drifting Type II bursts and broad-band Type IV bursts as they provide clue to predict space weather \citep{Kundu1965, McLean1985, Pick2004}.\\\\
Type III bursts are the most intense, frequently observed and fast drifting bursts in the solar corona and IPM. 
Most of the time, they occur in association with X ray and / or $H_{\alpha}$ flares \citep{Cane1988,White2007}. 
Type III bursts occur as isolated bursts that lasts in 1 - 3s, in groups that last in 10 minutes and as storms that lasts in few hours. 
The flare accelerated electrons that are travelling along open magnetic field lines setup the plasma oscillations (also known as Langmuir 
waves) during their passage in the corona and IPM and subsequent conversion of those oscillations into electromagnetic waves produces the Type III bursts \citep{Ginzburg1958,Zheleznyakov1970,Melrose1980, Mercier1975,Pick1986, Sasikumar2013, Dayal2019, Mah2020, Reid2014, Saint_Hilaire_2012}. 
In this article, we have carried out a statistical study of Type III bursts.\\\\
Type II bursts were discovered by \citet{Pay1947} and they are generated in association with CMEs that are moving at super-Alfvenic speeds \citep{Nelson1985,Cliver1999,Nindos2008,Nindos2011,Vrsnak2008}. Type II bursts are the signatures of particle acceleration caused by shock waves in the solar corona and IPM \citep{Gop2019}. Type IIs are slow drifting ($\rm \approx 1~MHz~s^{-1}$) bursts and it is 
widely accepted that they are originated due to the plasma emission mechanism \citep{McLean1985,Nelson1985}. 
Type II bursts often show fundamental and harmonic components with a frequency ratio $\approx 1:2$, and some times these components show band-splitting \citep{Vrs2001, Vas2014, Har2014, Har2015, Kis2016}. Type IV bursts are broadband quasi-continuum emissions that are mostly 
associated with the flares/CMEs \citep{Boi1957, Ste1982, Gary1985, Ger1986, Gop1987, Sas2014, Car2017}. 
Type IV bursts are further divided into two sub-categories: (i) stationary type IV bursts that occur during impulsive phase of the flares and the radio source location remains at same height, and (ii) moving type IV bursts which generally associated with CMEs. Imaging observations of moving Type IV bursts confirm that location of a radio source moves outward in the solar corona with speed of $\rm \approx 200 - 1500~ km~ s^{-1}$ \citep{Sas2014, Vas2016, Liu2018, Vas2019}. Their origin can be due to either plasma emission or gyro 
synchrotron emission mechanism \citep{Sas2014,Car2017}. Both Type II and Type IV bursts show a significant correlation with space weather
hazards \citep{White2007, Vor2020}. \\\\
Type II, Type III, and Type IV bursts provide clues to predict the space weather hazards because: (i) Type III bursts are triggered by 
solar flares, (ii) Type II burst originate from shocks observed in the solar corona and IPM, and (iii) moving Type IV bursts are 
associated with the core of the CMEs \citep{Sas2014}. It is worth mentioning that enhancement of X ray and EUV radiation due to the 
flares changes the conditions of the ionosphere and increases total electron content (TEC). Note that TEC of the ionosphere is a crucial 
parameter related to the frequency of radio waves which experience transmission or reflection from the ionosphere \citep{Car2020, Sel2015}. 
The CMEs and co-rotating interaction regions (CIRs) are responsible for geomagnetic storms and increased activity in the ionosphere. 
Note that they impact the satellite communication, telecommunication, Global Navigation Satellite Systems (GNSS). 
Also, the solar energetic particles damage the satellites and the radiation is hazardous for the astronauts and crew of the flights. \\\\  
We make a note that X ray and EUV flux and energetic particles from the solar flares influence the ionosphere within hours and impacts 
the HF communication whereas for CMEs to reach Earth takes 1 - 5 days depending on their speed and direction of arrival. 
The observations shown in this article suggests that the radio observations carried out using ground based e-CALLISTO network can be 
used to predict the space weather hazards.\\\\
A Compound Astronomical Low frequency Low cost Instrument for Spectroscopy and Transportable Observatory 
(CALLISTO; \url{http://www.e-callisto.org/}) is a radio spectrometer designed to monitor the radio transient emissions 24 hours a day 
\citep{Benz2009,Zuc2012, Sas2018}.
It is designed to operate in the frequency range 10 MHz -- 870 MHz and this frequency range probe the solar corona in the heliocentric 
distance range $\rm \approx 1 - 3~R_\odot$ \citep{Poh2007}. 
We make a note that there are $> 152$ stations operating around the globe and form an e-CALLISTO network. 
Presently about 52 of them regularly provide data to a server (1 frame in every 15 minutes) located at the University of Applied Sciences
(FHNW) in Brugg/Windisch, Switzerland. 
Therefore, routine detection of Type II, Type III and Type IV bursts along with other space based observations 
(like GOES X-ray flux, Extreme Ultraviolet, and white light coronagraph images) are useful for space weather forecasting 
agencies \citep{Prieto2020}. \\\\
This article is organized as follows. 
In Section 2, we describe the data used in this article. Section 3 describes a statistical study of Type III bursts 
(both isolated and group of Type III burst). 
This section also deals with the space weather implication of Type II, Type III and Type IV bursts.
The summary and conclusion of the article are presented in Section 4.
\section{Observation}
\begin{figure}[!ht]
 \centering
  \includegraphics[scale=.43]{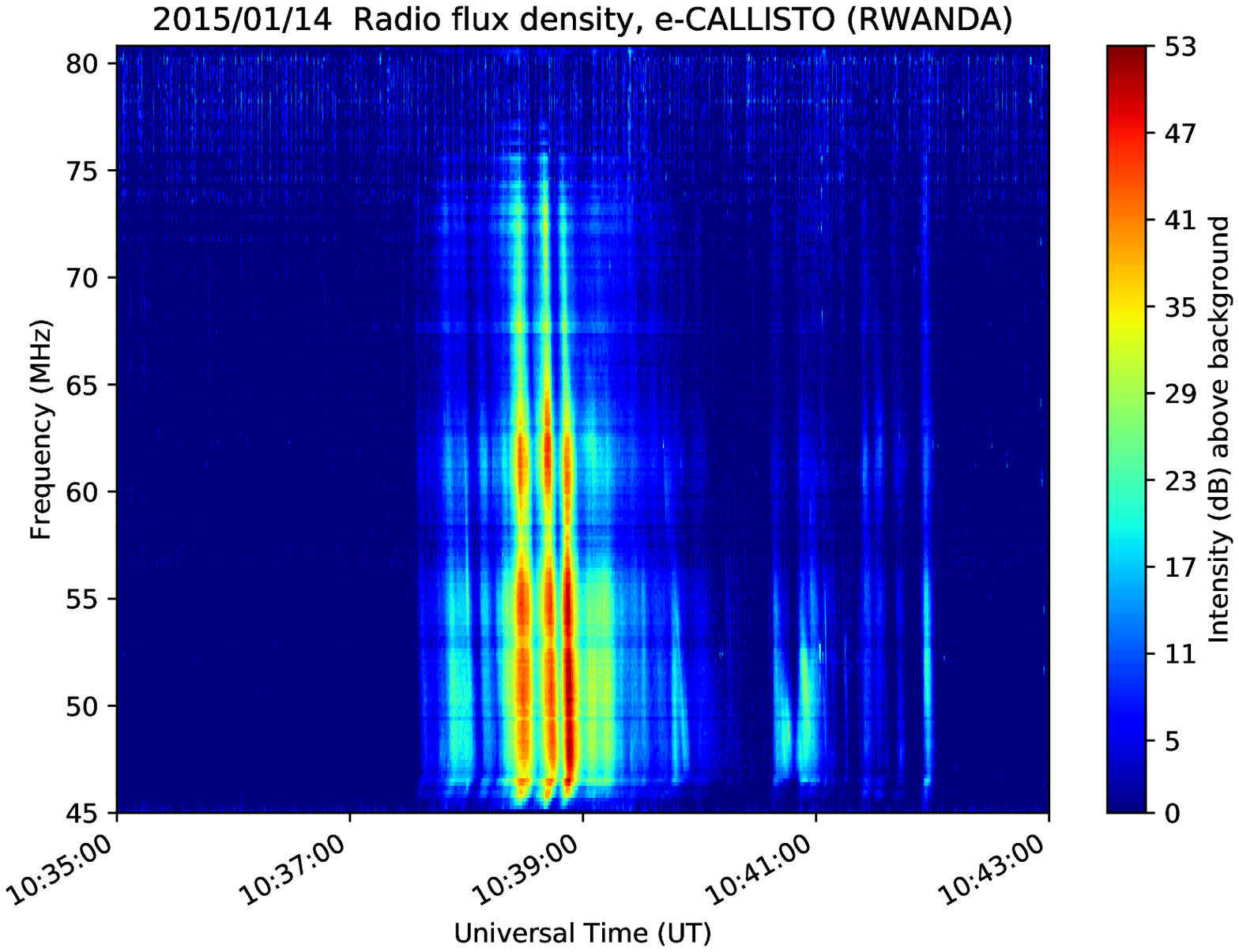} 
  \includegraphics[scale=.43]{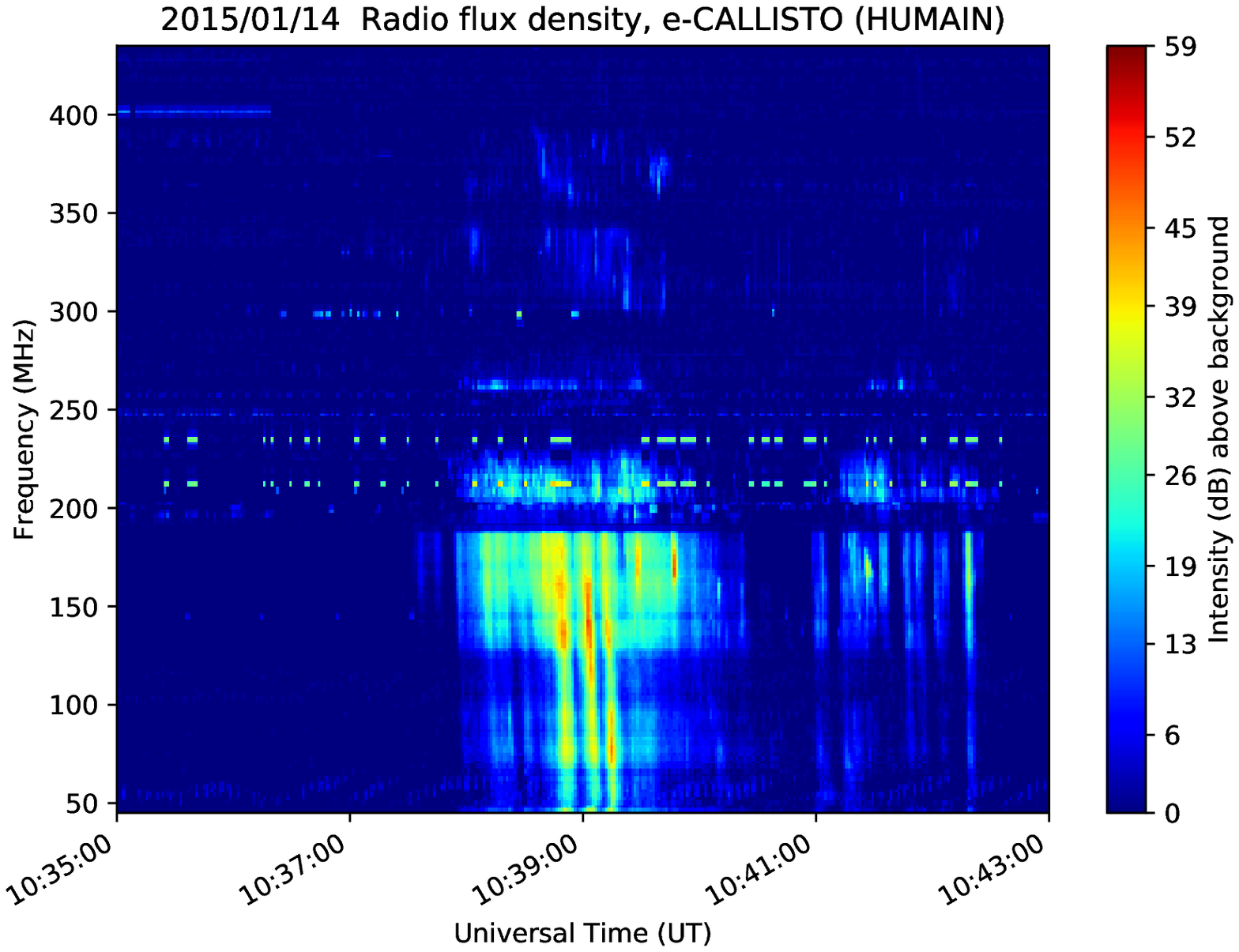}
\caption{The dynamic spectrogram shows group of Type III bursts observed on 14 January 2015. 
The left and right panels show the spectrograms observed using a CALLISTO spectrometer located at Kigali, Rwanda (45 - 80 MHz) and Royal 
Observatory of Belgium (45 - 437 MHz), respectively. This event is associated with a flare of class C1.9.}
\label{Fig1a}
\end{figure}

\noindent In this study, we used the dynamic spectrograms observed using e-CALLISTO network \citep{Benz2005,Benz2009}. 
Different stations operate over different frequency bands depending on the radio frequency interference (RFIs) and instrumental 
limitations. 
To begin with, we have prepared a list of Type III bursts using an online catalog ({\url{ftp://ftp.swpc.noaa.gov/pub/warehouse/}}) 
during 2010 - 2017 and manually cross-checked the bursts with the help of quick looks of the dynamic spectrograms that are available at 
e-CALLISTO website  ({\url{http://soleil.i4ds.ch/solarradio/callistoQuicklooks/}}). 
To cover the observations of 24 hours, we used data from different CALLISTO spectrometers that are listed in the Table \ref{Table1}. 
Different columns of Table \ref{Table1} indicate the station-id, hosted country, corresponding longitude and latitude of the station, and the range of operating frequency, respectively. \\\\
\begin{table}[!htb]
\centering
 \caption{Description of CALLISTO Spectrometers used in this work. Note that a burst can be observed by more than one spectrometer}
 \label{Table1}
\begin{tabular}{|c|l| l | c| c|c|}
\hline \bf{SN} & \bf{File ID} & \bf{Country} & \bf{Lat ($^\circ$)} & \bf{Long ($^\circ$)} & Frequency range (MHz) \\
\hline 1 & BLEN7M     & Switzerland   & 46.94  &  7.45  & 170--870\\
\hline 2 & BIR        & Ireland       &  53.09 & -7.90  & 10--400 \\
\hline 3 & GAURI      & India         &  16.61 &  77.51 & 45--410 \\
\hline 4 & GLASGOW    & UK            & 55.90  & -4.30  & 45--80.9 \\
\hline 5 & GREENLAND  &  Greenland    & 67.00  & -50.72 & 10--110 \\
\hline 6 & BLENSW     & Switzerland   & 47.34  &  8.11  & 10 --90 \\
\hline 7 & ROSWELL-NM & New Mexico    & 33.39  & -104.52& 15--80.9\\
\hline 8 & SSRT       & Siberia       &  6.43  &  2.88  & 45--450\\
\hline 9 & KRIM       & Ukraine       &  44.95 &  34.10 & 250--350\\
\hline 10 & OOTY       & India        & 11.41  &  76.69 & 45--450 \\
\hline 11 & HUMAIN     & Belgium      & 50.20  &  5.25  & 45--410 \\
\hline 12 & ALMATY     & Kazakhstan   &  43.22 &  76.85 & 45--400\\
\hline 13 & MRT        & Mauritius    & -20.16 &  57.74 & 45--450\\
\hline 14 & RWANDA     & Rwanda       &  -1.94 &  30.06 & 45--80.9 \\
\hline 15 & ESSEN      & Germany      &  51.39 &  6.97  & 20--80.9  \\
\hline 16 & OSRA       & Czech Rep.   &  49.90 & 14.78  & 150--870\\
\hline 17 & TRIEST     & Italy        &  45.64 & 13.77  & 220--445\\
\hline 18 & ALASKA     & Alaska       &  64.20 & -149.49&  210--450\\
\hline 19 & DENMARK    & Denmark      &  55.67 &  12.56 & 45--100 \\
\hline 20 & eC71       & Switzerland  & 47.24  & 8.92   & 48--90 \\
\hline 21 & KASI       & South Korea  & 36.35  & 127.38 & 45--450 \\
\hline 22 & PERTH      & Indonesia    & -31.65 & 115.86 & 45--870\\
\hline 23 & DARO       & Germany      &  51.77 & 6.62   & 30--90 \\
\hline 24 & Heiterswil-CH& Switzerland & 47.30 & 9.13   & 45--82\\
\hline 25 & INDONESIA  & Indonesia    & -6.84  & 107.93 & 45-80.9 \\
\hline 26 & ROZTOKY    & Slovakia     &  49.40 & 21.48  & 45--400 \\
\hline
\end{tabular}
\end{table}
In Figure \ref{Fig1a}, left and right panels show the dynamic spectrogram of group of Type III bursts (henceforth Type IIIg 
bursts) observed using a CALLISTO spectrometer located at Kigali, Rwanda (observed at 45 - 80 MHz) and at Royal Observatory of Belgium 
(ROB; observed at 45 - 437 MHz), respectively. We make a note that since the observations at Rwanda are carried out over a narrow
bandwidth when compared it with the ROB, the frequency resolution of former instrument is better than the latter and thus the left 
panel shows an intense and finer features. However, ROB observes the weak features of the burst beyond 80 MHz.\\\\
For all the bursts, we measured the start and stop time, and lower ($F_{L}$) and upper ($F_{U}$) frequencies manually. 
Although, different radio stations are operated over different range of operating frequencies (see Table \ref{Table1}), we selected $F_{L}$ 
and $F_{U}$ to be the lowest and highest frequencies among the simultaneous observations carried out by different stations, respectively.
Since Type III bursts are triggered by solar flares and some of the flares are accompanied by CMEs, we have carried out a statistical 
analysis of them. The details of solar flares and CMEs and the associated radio bursts are taken from Heliophysics Event Catalogue 
({\url{http://hec.helio-vo.eu/hec/hec\_gui.php}}). 
If a radio burst is present between onset and end times of the flare then we treat that to be flare associated. 
Otherwise, the burst is treated as non-flare associated. 
\section{Results and Discussions}
\subsection{Statistical study of Type III bursts}
Many authors have attempted to study a correlation between occurrence of the total number of type III bursts and Sunspot number (SSN)
\citep{Lobzin2011, Huang2018, Mah2020}. Using the L-band observations, \citet{Huang2018} have studied 2384 SRBs observed 
during 1997 - 2016 and reported that occurrence of SRBs closely track the solar cycle. 
In the article, we have identified 12971 Type III bursts (during 2010-2017) that are reported by Space Weather Prediction Center (SWPC) 
of National Oceanic and Atmospheric Administration (NOAA) and explored their relationship with the solar cycle.  
\begin{figure}[!ht]
\centering
\includegraphics[scale=.99]{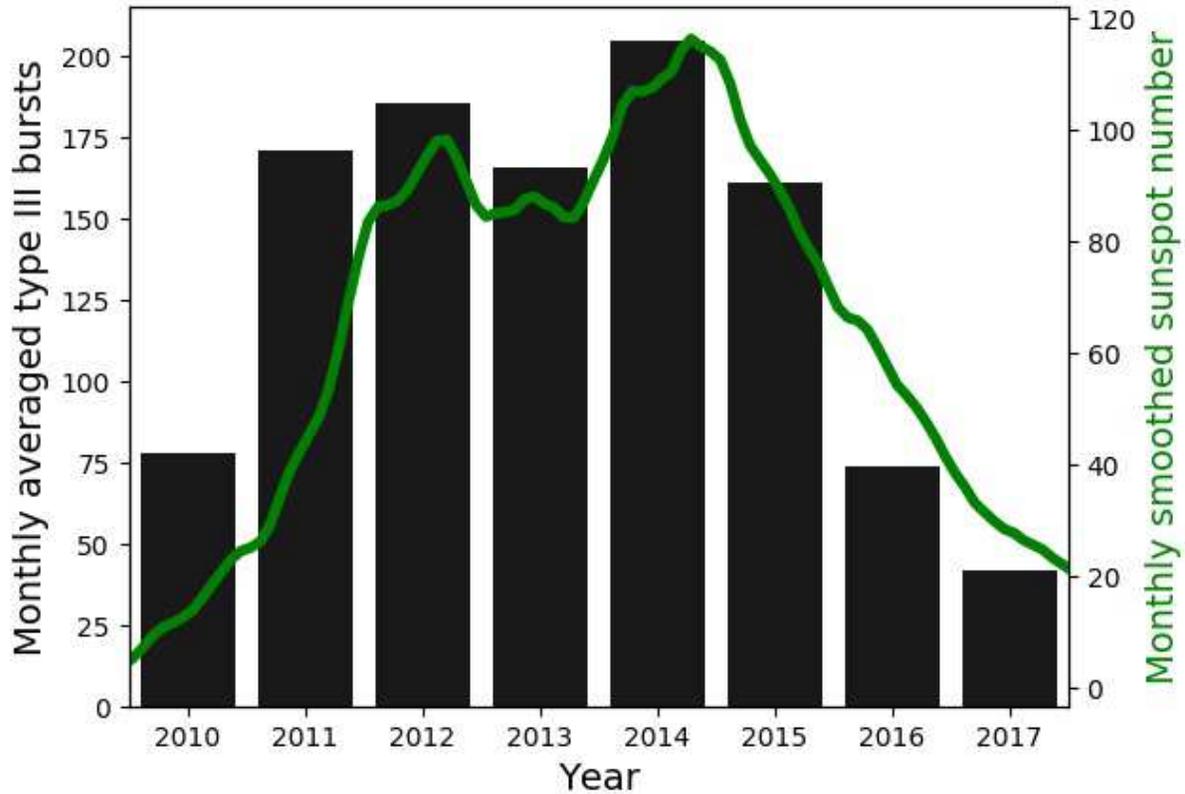}
\caption{Occurrence of Type III bursts and the solar activity. Occurrence rate of Type III bursts is well correlated with two peaks of 
the Sunspots observed during solar cycle 24.}
\label{Fig2}
\end{figure}
Figure \ref{Fig2} shows the monthly averaged number of Type III bursts (i.e., number of Type III bursts observed in a year / 12) 
observed in different years. The green curve shows the revised version of the Sunspot number that are available at {\url{http://www.sidc.be/silso/datafiles}} \citep{Clette2016}. 
From Figure \ref{Fig2}, it is evident that occurrence of Type III bursts highly correlates with the solar cycle. 
It is noteworthy that number of Type III bursts follow the two peaks of Sunspots that occurred during maximum period of solar cycle 24 
(SC24; i.e., in the years 2012 and 2014). Also, a fewer number of bursts that are observed during the solar minimum 
(i.e. in the years 2010 and 2017) are due to the lower solar activity.\\\\

\begin{figure}[h!]
\centering
 \includegraphics[scale=.5]{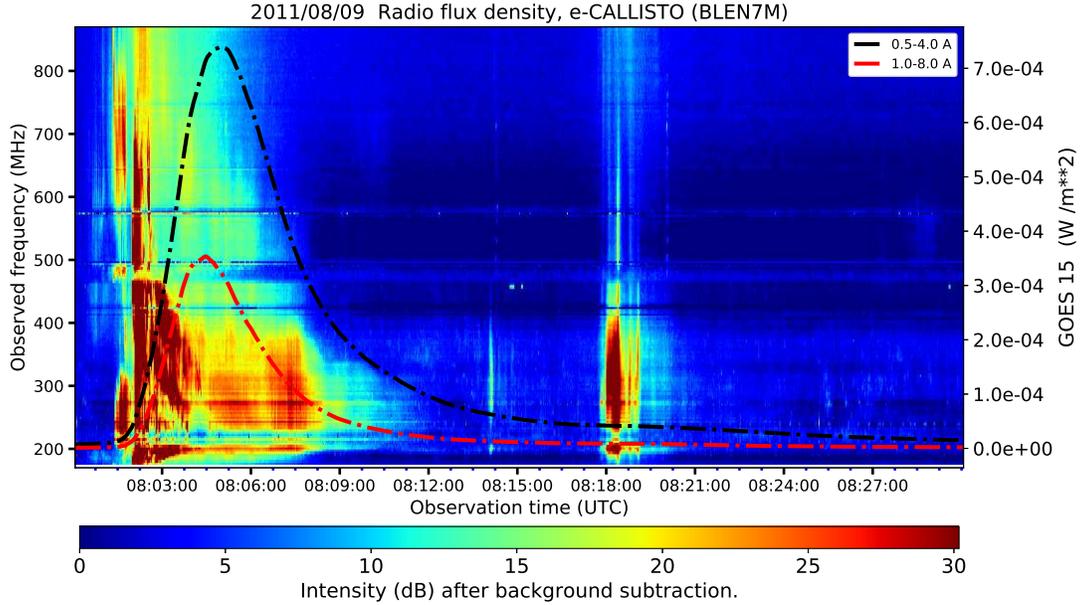}
 \caption{Dynamic spectrogram show two groups of Type III bursts. The one at 08:01 UT occurred during the ascending phase of X6.9 
 solar flare. Type IIIg that occurred at 08:17 UT is triggered by the flare during its decay phase. 
 The red and black curves shows the GOES X ray flux.}
 \label{Fig3}
\end{figure}

\begin{figure}[ht!]
\centering
\includegraphics[scale=.99]{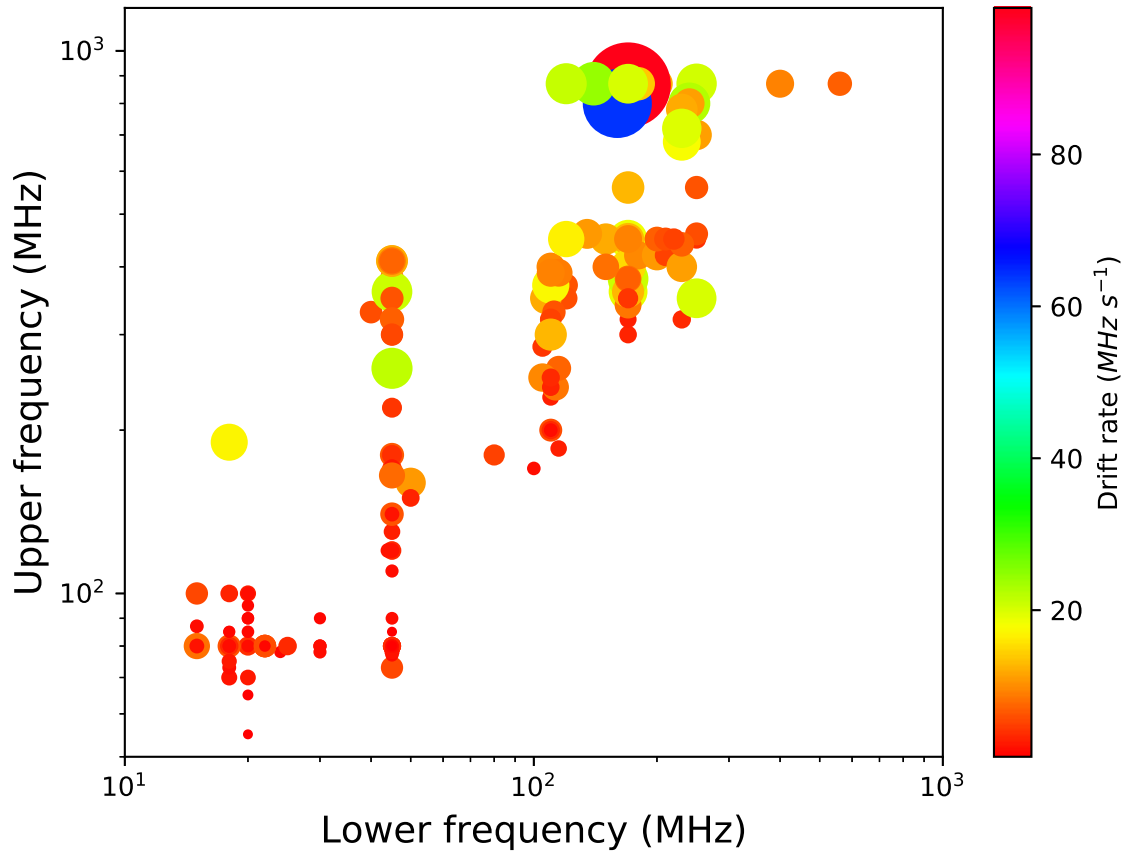}
\caption{Distribution of drift rates of isolated Type III bursts measured for a given lower and upper frequencies. 
Note that larger size of the marker indicate higher drift rates.}
\label{Fig4}
\end{figure}

Among the 12971 Type III bursts, we have randomly selected 619 intense type III bursts by visual inspection (including 221 isolated Type III and 398 Type IIIg bursts) and carried out a statistical study. We found that out of 619 Type III bursts, $\approx$65\% bursts are associated with soft and/or $H_\alpha$ flares while $\approx$ 45\% of 
them are accompanied by CMEs (see Table \ref{tab:summary}). 
A detailed table of type III bursts (that are used for statistical study) can be accessed from {\url{http://www.e-callisto.org/GeneralDocuments/Supplementary_material.pdf}}. 
The remaining non-flare associated bursts presumably originated due to the weak energy releases that are present in the solar 
corona \citep{Ramesh2010,Ramesh2013,Saint_Hilaire_2012,Sasikumar2013,Mugundhan2017,James2017,James2018,Sharma2018, Mah2020}. 
Earlier \citet{Dayal2019} reported that out of 238 Type III bursts that are observed during $\approx$ 02:30 UT -- 11:30 UT in 2014, 
88 bursts are associated with the GOES X ray flares / $H_\alpha$ flares.
\citet{Mah2020} reported that out of 1531 type III bursts, 426 bursts are associated with GOES X ray / $H_\alpha$ flares. 
The non-flare associated Type III bursts may be triggered by  $H_{\alpha}$ ejecta, X-ray bright points, soft X-ray transient brightening,
and soft X-ray or/and extreme-UV (EUV) jets etc \citep[more references therein]{Alissandrakis2015}. 
\begin{table}[h!]
\caption{Summary of observations}.             
\label{tab:summary}     
\centering                          
\begin{tabular}{lc lc lc}       
\hline\hline                 
      & Isolated & Group of & Isolated + Group of   \\  
Parameter      & Type III bursts & Type III bursts & Type III bursts   \\  

\hline
Studied Type III bursts & 221 & 398 & 619 \\
Flare associated        & 135 & 267 & 402\\
CME accompanied          & 90 & 192 & 288 \\
Flare associated (\%)   & 61 & 67 & 65\\
CME accompanied (\%)     & 40 & 48 & 45 \\
\hline                                   
\end{tabular}
\end{table}\\
Further, we have divided the total bursts into two categories and studied their characteristics separately. 
The first category comprises a total of 221 isolated type III bursts (with duration $<$ 60 seconds) and the second category comprises 
of 398 groups of type III bursts (with duration $\approx$ 1 min -- 10 min).
Figure \ref{Fig3} shows a CALLISTO spectrogram obtained at Bleien, Switzerland of an intense Type IIIg burst at the early 
phase of flare of class X6.9 which is followed by another Type IIIg burst occurred during its decay phase. 
Note that Type IIIg burst observed at 08:05 UT is associated with an ascending phase of flare while the burst at 08:17 UT is 
associated with decay phase of the flare. The GOES X ray light curves are shown in black and red colors in Figure \ref{Fig3}.
\\
For all the isolated Type III bursts we measured the start time ($t_i$), stop time ($t_f$), upper frequency ($F_U$) and lower frequency 
($F_L$) of the bursts and measured their drift rates using, 
\begin{equation}
 \frac{df}{dt}=\left|\frac{F_{U}-F_{L}}{t_f-t_i}\right|
 \label{ED}
\end{equation} 
Figure \ref{Fig4} illustrates the distribution of drift rates for different lower and upper frequencies. 
From the plot it is found that at lower frequencies drift rates are smaller compared with the higher frequencies. 
One possible explanation is, when the flare accelerated electrons travel along the open magnetic field lines which diverges with the 
radially increasing distance, duration of the radio bursts increases and thus lower drift rates are expected at lower frequency 
(i.e., in outer layers of corona).
\citet{Reid2014} interpret that faster drift rates in the deep solar corona is due to rapid change of frequency of the background 
plasma with the increasing distance.

In this study, we have found that drift rates of the isolated bursts vary in the range $\rm 0.70 - 99~ MHz~s^{-1}$. 
Earlier \citet{Zhang2018} have studied 1389 simple isolated Type III bursts observed using the Nan\c{c}ay decameter array over the 
frequecy range 10 MHz - 80 MHz. 
The authors have reported that drift rates of the Type III bursts range from $\rm 2 ~MHz~s^{-1}$ to $\rm 16~MHz~s^{-1}$, with a median 
value of $\rm 6.94~ MHz~s^{-1}$. In Figure \ref{Fig4}, both color and the size of the marker indicate different drift rates for a given lower and upper frequency. 

\begin{figure}[h!]
 \centering
  \includegraphics[scale=.55]{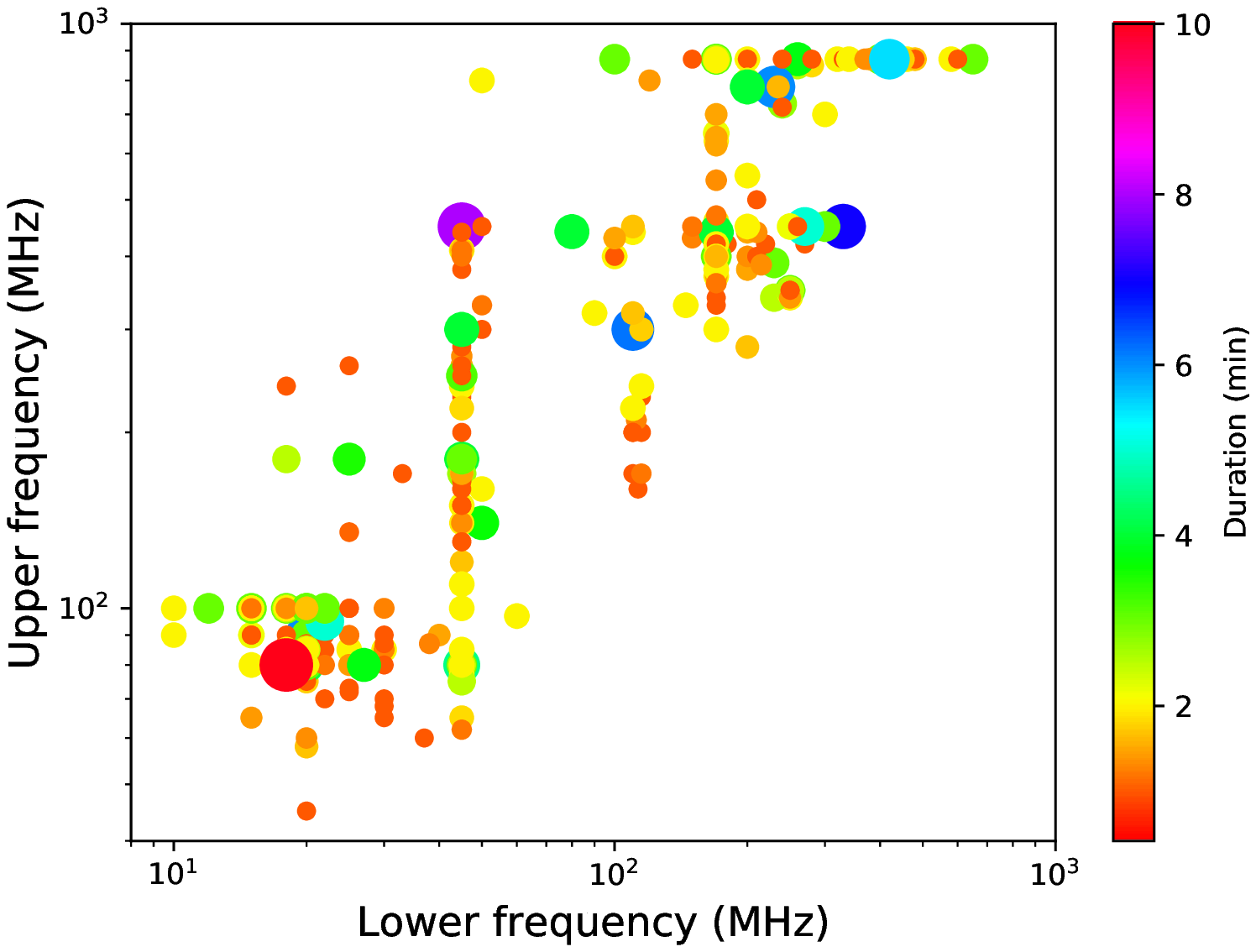}
  \includegraphics[scale=.55]{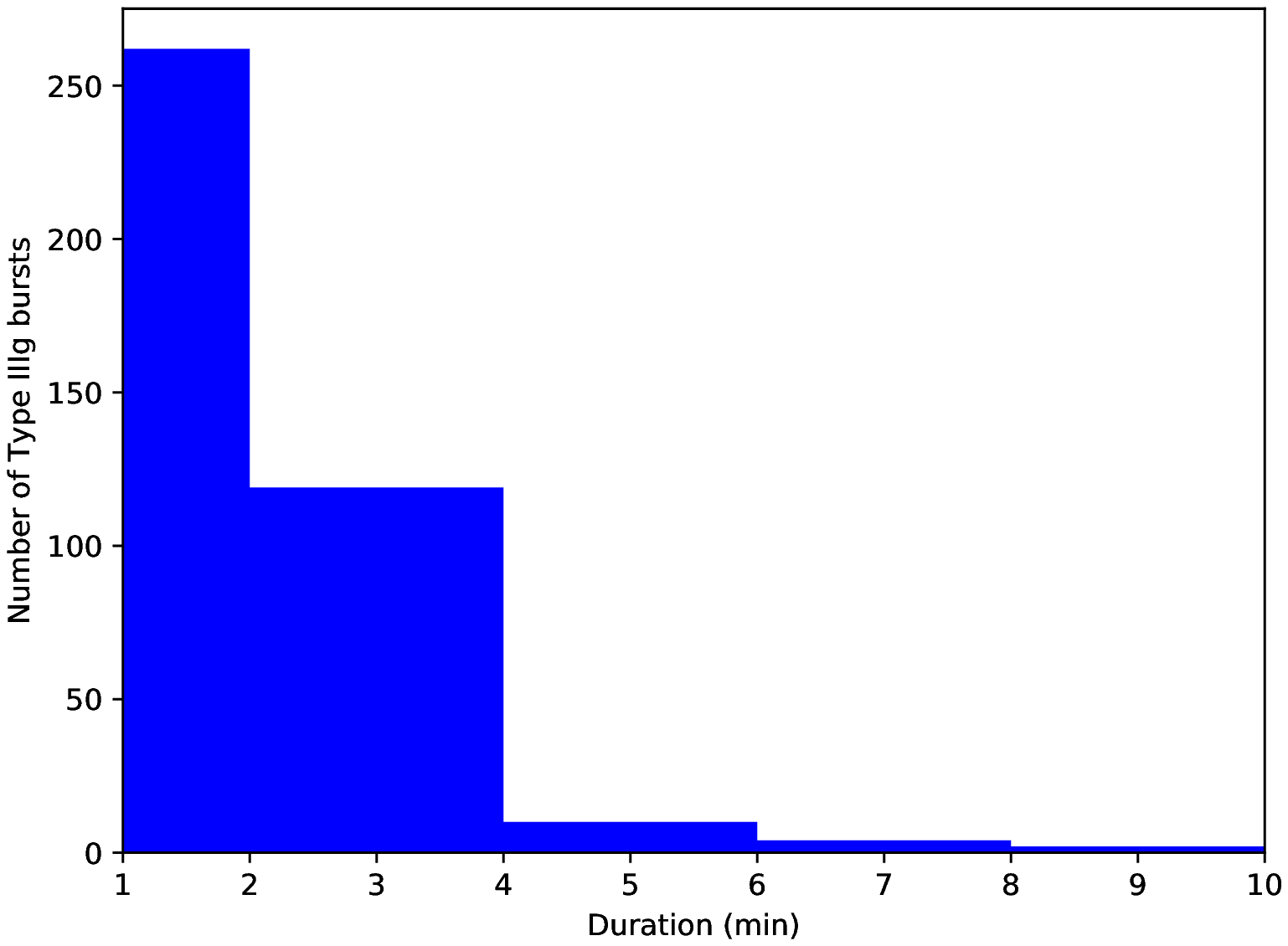}
\caption{The left panel shows the time duration over which Type IIIg bursts are lasted. The color and size of the marker indicate 
the time interval over which they lasted. Note that larger the size of the marker means longer duration. The right panel shows number of 
Type III bursts vs the time interval over which they are observed.}
\label{Fig5}
\end{figure}

We have studied the duration (i.e. stop time minus start time) of 398 Type IIIg bursts. 
The left panel of Figure \ref{Fig5} shows the way duration of Type IIIg bursts are varied. 
It is also evident that duration of Type IIIg bursts are independent of lower and upper frequency cut-offs. 
Different colors and sizes of the markers indicate duration of Type IIIg bursts for a given lower and upper frequencies. 
The histogram in right panel shows that 388 Type IIIg bursts (i.e. 97.5 \%) lasted within 4 minutes of time.
\subsection{Space weather implication}
\noindent In this article, we investigated two events that caused the radio blackouts in different parts of Earth. 
Using those events we attempt to explain the way radio emissions can help to predict the space weather hazards. 
\subsubsection{The 6th September 2017 event}
\begin{figure}[h!]
\centering
\includegraphics[scale=.5]{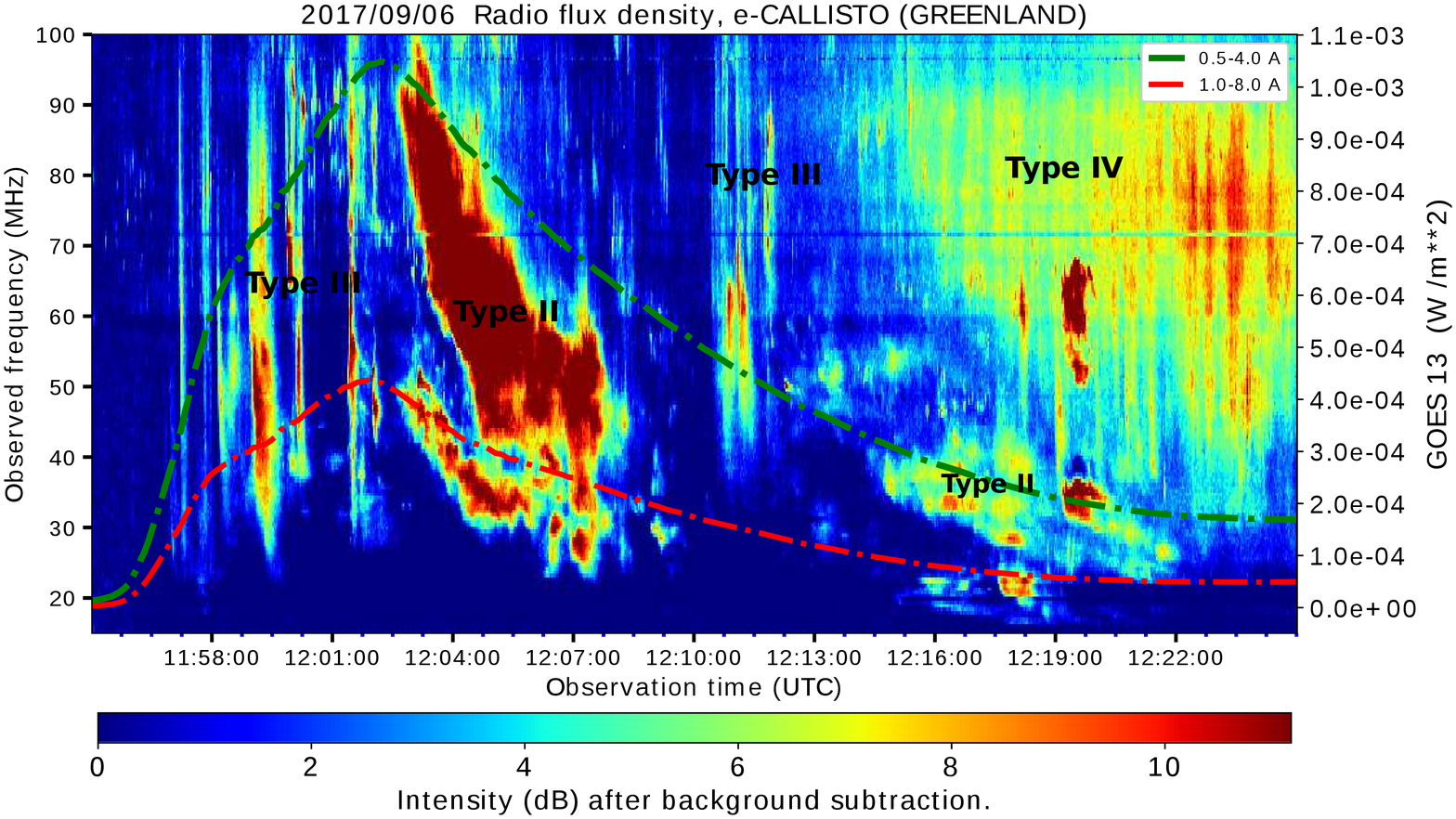}
\includegraphics[width=0.7\textwidth]{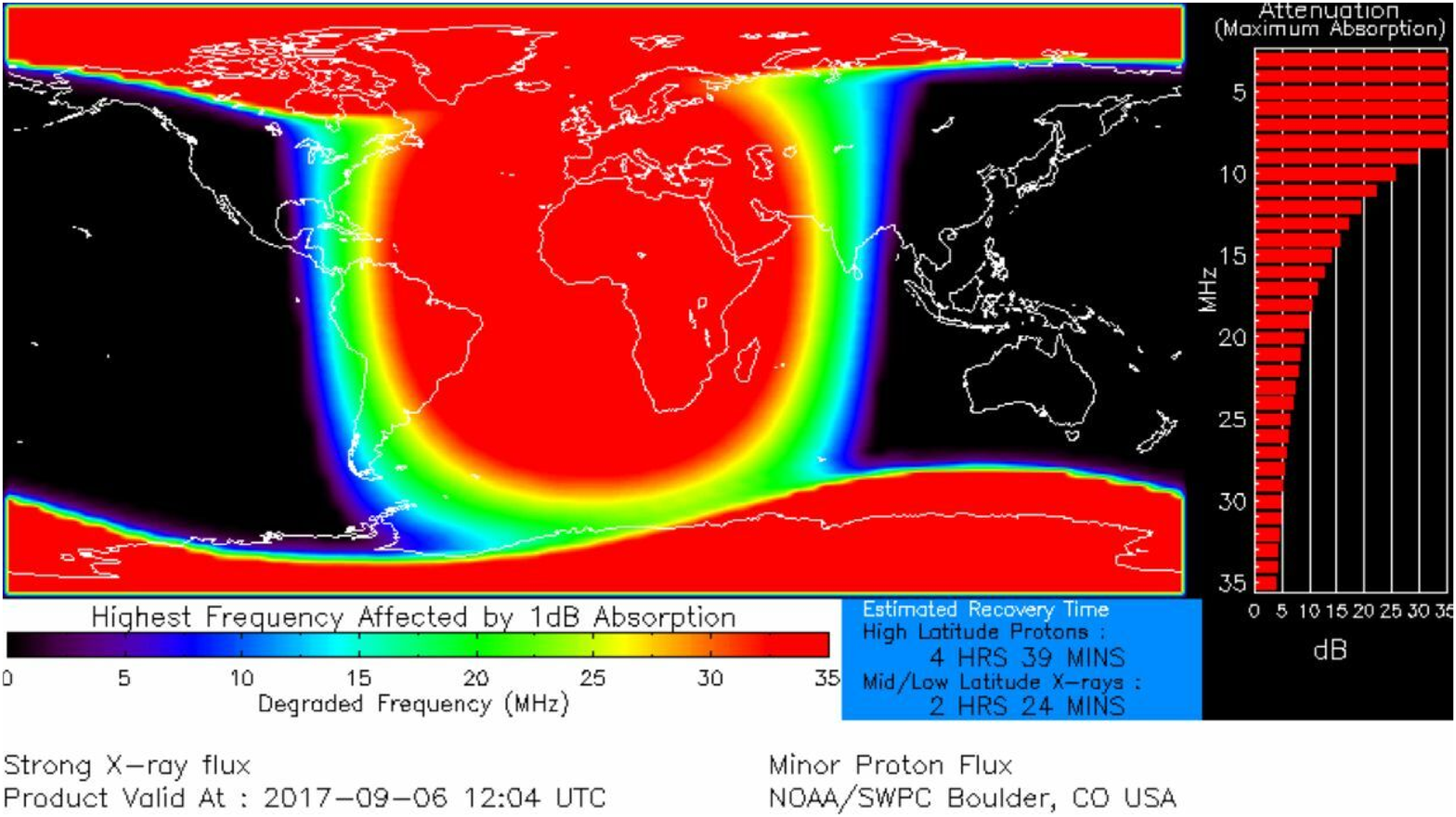}
\caption{The top panel shows radio emissions that are observed on 6 September 2017. The red and green curves show a GOES X ray flux 
during the radio observations. 
It shows Type IIIg burst during rising and decay phase of the X ray flare and are followed by two Type II and a stationary Type IV burst. 
Lower panel shows a radio blackout that occurred over Europe, Africa and the Atlantic Ocean on 6 September 2017 due to the enhancement of 
X-ray flux and UV radiation at the Earth. Credited to SWPC.}
  \label{Fig6}
  \end{figure}

On 6 September 2017, an intense and large group of Type III burst is observed at Greenland station, and its emission coincided with the
early phase of large flare of class X9.3 that peaked at 12:02 UT as shown in Figure \ref{Fig6}. 
The explosion of this flare was stronger and the most intense in SC24. The eruption is caused by a complex Active Region (AR) 12673 
located at S09W34. The green and red curves on Figure \ref{Fig6} are the GOES X-ray light curves. \\\\
The group of Type III bursts started at 11:57 UT and disappeared at 12:12 UT in the frequency range of 18 MHz -- 100 MHz. 
Type IIIg bursts are observed during both ascend and decay phase of the flare. 
Following Type IIIg, two Type II bursts and a stationary Type IV burst are observed. The first Type II burst is observed from 
12:02:37 to 12:09:31 UT and in the frequency range 18 MHz - 100 MHz. The measured drift rate of the burst is $\approx 0.2~  MHz~s^{-1}$. 
The second Type II burst is observed between 12:13:02-12:21:57 UT and in the frequency range 18 MHz - 60 MHz. 
The measured drift rate is $\approx 0.1~ MHz~s^{-1}$. Another interesting observation is both Type II bursts show the fundamental and 
harmonic emissions. Also, a Type IV burst is observed at 12:14 UT. \\\\
Here, we would like to emphasize the following points: (i) Type IIIg bursts are triggered by flare accelerated electrons, (ii) both 
type II bursts are triggered by CME shocks, and (iii) Type IV burst is associated with CME. In other words, the radio bursts that 
are observed in the Figure \ref{Fig6} are basically signatures of flares and/or CMEs. 
Therefore, such observations can be used to forecast the space weather even in the absence of white light coronagraph observations. 
Also, since there are $> 152$ CALLISTO stations are operating around the globe to monitor the transient emissions from the solar corona, 
e-CALLISTO network is a powerful tool to forecast the space weather hazards.   
Note that the CMEs associated with the radio emission shown in Figure \ref{Fig6} is Earth-directed and a nearly symmetrical halo with an 
estimated sky-plane velocity of 1571 km s$^{-1}$ as observed by SOHO coronagraph at 12:24 UT. 
Also, the CME caused a significant compression to the day side Earth's magnetosphere and prompted a severe (G4) geomagnetic storm 
($\rm K_{pmax}= 8.3$ and $\rm Dst_{min}= -124$ nT) that reached the Earth on 7-- 8 September 2017 (see Figure \ref{Fig6}).

Furthermore, intense radio emissions shown in Figure \ref{Fig6} are associated with major impulsive increases with X-rays and EUV 
emission. 
\citet{Sato2019} have reported the impact of the 6 September 2017 on GNSS signals and inferred that the bursts are strong enough to give 
rise sudden disturbances on space weather environment.
The EUV emission is caused a prompt ionization enhancement in the Earth's upper atmosphere. Also the energetic particles from the Sun 
arrived at the Earth within few hours after the flare and resulted a large enhancement of high energy proton levels and caused a shortwave
radio blackout over Europe, Africa and the Atlantic Ocean as shown in lower panel of Figure \ref{Fig6}.
Moreover, a rapid and comprehensive ionization of the equatorial upper atmosphere has disrupted HF communications 
while emergency managers were struggling to provide critical recovery services (e.g., \citealp{NCEI2017}). 
Some of these issues were reported by the Hurricane Weather Net (HWN), and the French Civil Aviation Authority (DGAC) \citep{Redmon2018}.
\subsubsection{The 7th March 2010 event}
\begin{figure}[h!]
\centering
\includegraphics[scale=.5]{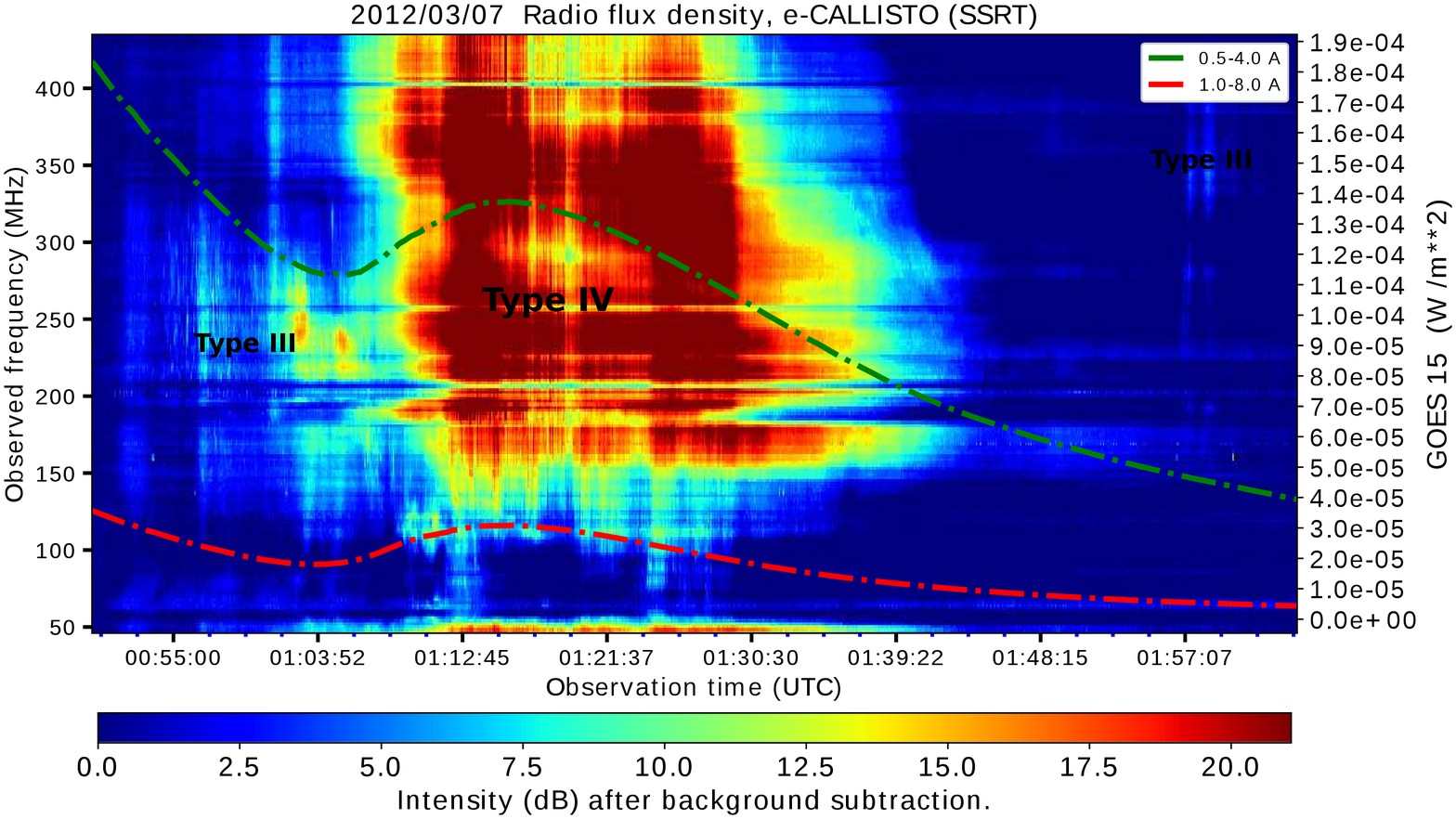}
\includegraphics[width=0.7\textwidth]{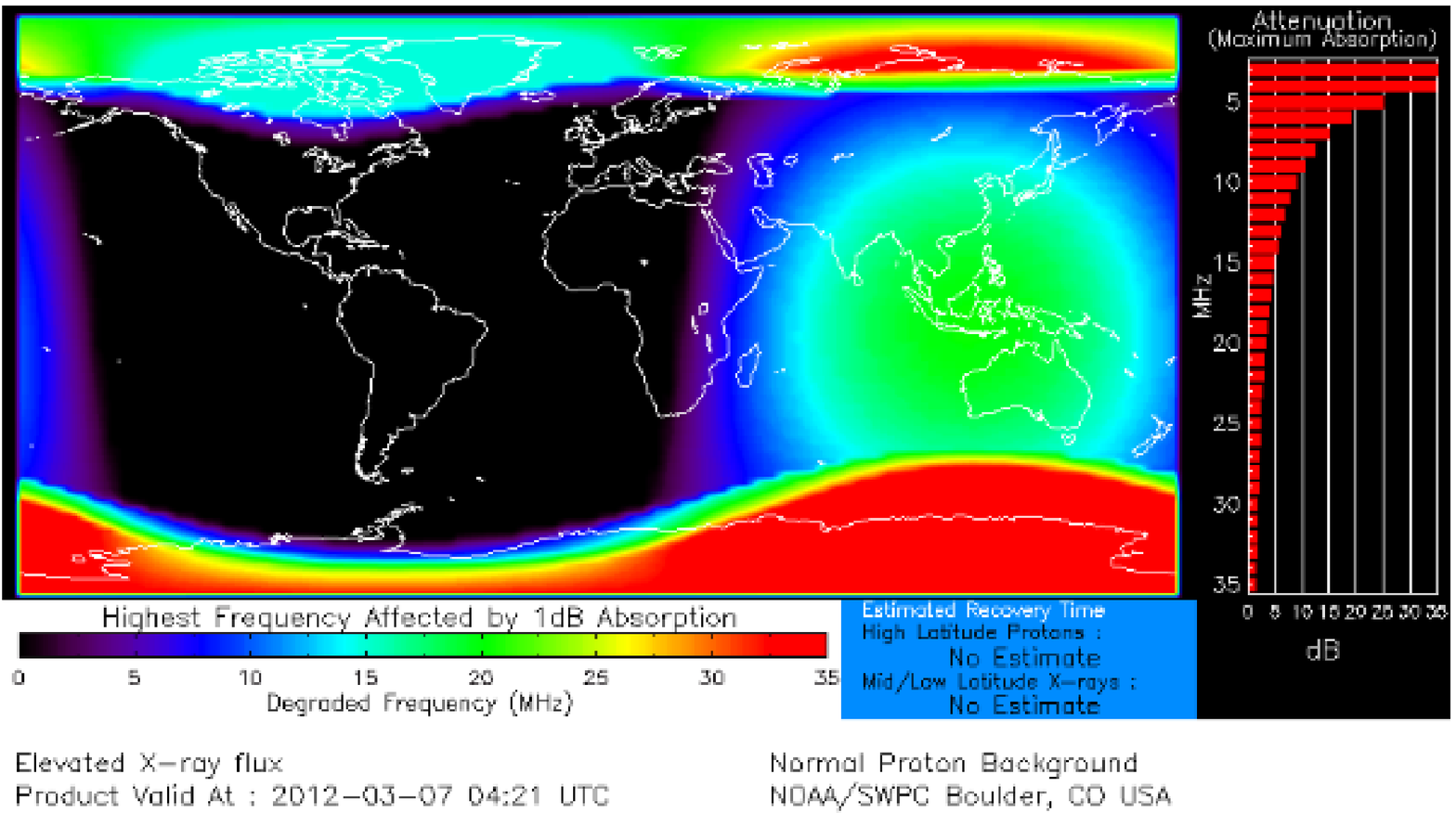}

\caption{The top panel shows radio emissions observed on 7 March 2012 by Badary station in Russia. 
The red and green curves show a GOES X ray flux during the radio observations. 
It shows Type IIIg bursts during rising and decay phase of the X ray flare and a stationary type IV burst. 
Lower panel shows a radio blackout that occurred over Asia, Australia, and Indian and Pacific Oceans on 7 March 2012 due to the 
enhancement of X-ray flux and UV radiation at the Earth. Credited to SWPC.}
  \label{Fig7}
  \end{figure}

Type IIIg burst and a stationary type IV burst shown in Figure \ref{Fig7} is observed by Badary observatory (SSRT site), Siberia, 
Russian Federation on 7 March 2012. 
These radio emissions are triggered by GOES X5.4 flare and its onset and peak times are 00:02 UT and 00:24 UT, respectively. 
This flare is associated with active region AR11429 located at N18E31. The second flare is triggered by GOES X1.3 class flare and its 
onset and peak times are 01:05 UT and 01:14 UT, respectively. This flare also associated with an active regions AR11429 located at N15E26.
The green and red curves in Figure \ref{Fig7} show the X ray light curves observed by GOES satellite.\\\\
Type IIIg burst observed at 00:56 UT in the frequency range of 90 MHz -- 320 MHz is followed with an intense Type IV burst. 
This Type IIIg is triggered by X5.4 class flare during its decay phase. Another Type IIIg burst occurred at 01:58 UT is triggered by X1.3 
flare during its decay phase. 
These flares are associated with two CMEs launched from AR11429 region. The estimated speeds of first and second CME are $\approx 2200$ 
and $\approx 1800~ km~s^{-1}$, respectively \citep{Patsourakos2016}.
This event caused radio blackouts in Asia and Australia, and Indian and Pacific Oceans on 07 March 2012 at 04:21 UT (Figure \ref{Fig7}) 
and is followed by the aurora. The main phase of the magnetic storm starts at $\approx$ 02:00 UT on 7 March 2012 and reached a maximum at 
$\approx$ 05:15 UT, with Dst$=$ -74 nT and $\rm K_{pmax}= 6.0$ \citep{Tsurutani2014}.
The ionospheric disturbances caused by this storm is described by \citet{Krypiak2019}.
\section{Summary and conclusion}
\noindent In this study, we report a statistical analysis of Type III radio bursts and their correlation with the Sunspot number. 
During solar cycle 24, we have identified 12971 Type III bursts (during 2010-2017) by making use of the reports of Space Weather 
Prediction Center (SWPC) of National Oceanic and Atmospheric Administration (NOAA). 
We found that the occurrence of Type III bursts well correlates with the Sunspot number. 
Further, we have randomly selected 619 Type III bursts comprising 221 isolated and 398 group of bursts and carried out a statistical study of 
Type III bursts using the dynamic spectrograms observed by e-CALLISTO network. We found that $65 \%$ of them are flare associated and 
$45\%$ of the bursts are accompanied by a CME. 
The remaining non-flare associated Type III bursts are believed to originate from weak energy release events in the solar atmosphere. 
We have also found that drift rates of isolated bursts vary in the range $\rm 0.70 - 99~ MHz~s^{-1}$. 
This study confirms that drift rates at high frequency are larger than the lower frequencies. 
It can be due to the following facts: (i) electrons that are travel along the open magnetic field lines diverge with the 
radially increasing distance, therefore duration of the radio bursts increases resulting lower drift rates, and (ii) 
faster drift rates in the deep solar corona is due to rapid change of frequency of the background 
plasma with the increasing distance \citep{Reid2014} .
We have studied the duration of Type IIIg bursts and found that $95.5\%$ of them are lasted within $4$ minutes and they are independent 
of lower and upper frequency cut-offs. \\

The meter wave solar radio bursts are launched in the same layer of the solar atmosphere where geo-effective disturbances initiate and 
hence, they can be potential signatures to forecast the space weather hazards. 
In this study we show two of such events that have caused radio blackouts near Earth. 
As previously mentioned, type III bursts are signatures of solar flares and Type II and Type IV bursts are the signatures of CMEs. The enhanced radiation at X ray and EUV wavelengths along with the energetic particles reach the Earth 
within few hours and influences the ionosphere and impacts the HF communications. 
Note that it takes 1-5 days for a CME to impact the earth depending on the speed and direction. 
Ground based e-CALLISTO network with more than 152 stations located at various longitudes is capable of observing solar radio transient 
emissions 24 hours a day. Furthermore, different types of radio bursts are associated with various solar transients and therefore, they can play a crucial role in predicting the space weather.

\acknowledgments
This work was supported by International Science Programme (ISP) through Rwanda Astrophysics, Space and Climate Science 
Research Group (RASCSRG).
We thank FHNW, Institute for Data Science in Brugg/Windisch, Switzerland for hosting the e-Callisto network; SOHO/LASCO; NOAA; GOES;
SWPC and WDC-SILSO, Royal Observatory of Belgium, Brussels to make their data available online.
The author (C. Monstein) thanks the ISSI - Bern International Team of ‘Why Ionospheric Dynamics and Structure Behave Differently in 
The African Sector’ (the team leaders E. Yizengaw \& K. Groves) for valuable discussions about part of the results that are included 
in this paper. 
We thank both the referees for providing useful and encouraging comments and suggestions which helped in improving the manuscript. 

\bibliographystyle{aasjournal}
\bibliography{main}

\begin{thebibliography}{}
\expandafter\ifx\csname natexlab\endcsname\relax\def\natexlab#1{#1}\fi
\providecommand{\url}[1]{\href{#1}{#1}}
\providecommand{\dodoi}[1]{doi:~\href{http://doi.org/#1}{\nolinkurl{#1}}}
\providecommand{\doeprint}[1]{\href{http://ascl.net/#1}{\nolinkurl{http://ascl.net/#1}}}
\providecommand{\doarXiv}[1]{\href{https://arxiv.org/abs/#1}{\nolinkurl{https://arxiv.org/abs/#1}}}

\bibitem[{{Alissandrakis} {et~al.}(2015){Alissandrakis}, {Nindos},
  {Patsourakos}, {Kontogeorgos}, \& {Tsitsipis}}]{Alissandrakis2015}
{Alissandrakis}, C.~E., {Nindos}, A., {Patsourakos}, S., {Kontogeorgos}, A., \&
  {Tsitsipis}, P. 2015, A\&A, 582, A52, \dodoi{10.1051/0004-6361/201526265}

\bibitem[{{Benz} {et~al.}(2005){Benz}, {Monstein}, \& {Meyer}}]{Benz2005}
{Benz}, A.~O., {Monstein}, C., \& {Meyer}, H. 2005, Sol. Phys., 226, 143,
  \dodoi{10.1007/s11207-005-5688-9}

\bibitem[{{Benz} {et~al.}(2009){Benz}, {Monstein}, {Meyer}, {Manoharan},
  {Ramesh}, {Altyntsev}, {Lara}, {Paez}, \& {Cho}}]{Benz2009}
{Benz}, A.~O., {Monstein}, C., {Meyer}, H., {et~al.} 2009, Earth Moon and
  Planets, 104, 277, \dodoi{10.1007/s11038-008-9267-6}

\bibitem[{{Boischot}(1957)}]{Boi1957}
{Boischot}, A. 1957, Academie des Sciences Paris Comptes Rendus, 244, 1326

\bibitem[{{Cane} \& {Reames}(1988)}]{Cane1988}
{Cane}, H.~V., \& {Reames}, D.~V. 1988, ApJ, 325, 901, \dodoi{10.1086/166061}

\bibitem[{{Carley} {et~al.}(2017){Carley}, {Vilmer}, {Sim{\~o}es}, \& {{\'O}
  Fearraigh}}]{Car2017}
{Carley}, E.~P., {Vilmer}, N., {Sim{\~o}es}, P. J.~A., \& {{\'O} Fearraigh}, B.
  2017, A\&A, 608, A137, \dodoi{10.1051/0004-6361/201731368}

\bibitem[{{Carley} {et~al.}(2020){Carley}, {Baldovin}, {Benthem}, {Bisi},
  {Fallows}, {Gallagher}, {Olberg}, {Rothkaehl}, {Vermeulen}, {Vilmer}, \&
  {Barnes}}]{Car2020}
{Carley}, E.~P., {Baldovin}, C., {Benthem}, P., {et~al.} 2020, Journal of Space
  Weather and Space Climate, 10, 7, \dodoi{10.1051/swsc/2020007}

\bibitem[{{Clette} {et~al.}(2016){Clette}, {Lef{\`e}vre}, {Cagnotti},
  {Cortesi}, \& {Bulling}}]{Clette2016}
{Clette}, F., {Lef{\`e}vre}, L., {Cagnotti}, M., {Cortesi}, S., \& {Bulling},
  A. 2016, Sol. Phys., 291, 2733, \dodoi{10.1007/s11207-016-0875-4}

\bibitem[{{Cliver} {et~al.}(1999){Cliver}, {Webb}, \& {Howard}}]{Cliver1999}
{Cliver}, E.~W., {Webb}, D.~F., \& {Howard}, R.~A. 1999, Sol. Phys., 187, 89,
  \dodoi{10.1023/A:1005115119661}

\bibitem[{{Gary} {et~al.}(1985){Gary}, {Dulk}, {House}, {Illing}, {Wagner}, \&
  {Mclean}}]{Gary1985}
{Gary}, D.~E., {Dulk}, G.~A., {House}, L.~L., {et~al.} 1985, A\&A, 152, 42

\bibitem[{{Gergely}(1986)}]{Ger1986}
{Gergely}, T.~E. 1986, Sol. Phys., 104, 175, \dodoi{10.1007/BF00159959}

\bibitem[{{Ginzburg} \& {Zhelezniakov}(1958)}]{Ginzburg1958}
{Ginzburg}, V.~L., \& {Zhelezniakov}, V.~V. 1958, Soviet Ast., 2, 653

\bibitem[{{Gopalswamy} \& {Kundu}(1987)}]{Gop1987}
{Gopalswamy}, N., \& {Kundu}, M.~R. 1987, Sol. Phys., 114, 347,
  \dodoi{10.1007/BF00167350}

\bibitem[{{Gopalswamy} {et~al.}(2019){Gopalswamy}, {M{\"a}kel{\"a}}, \&
  {Yashiro}}]{Gop2019}
{Gopalswamy}, N., {M{\"a}kel{\"a}}, P., \& {Yashiro}, S. 2019, Sun and
  Geosphere, 14, 111, \dodoi{10.31401/SunGeo.2019.02.03}

\bibitem[{{Hariharan} {et~al.}(2015){Hariharan}, {Ramesh}, \&
  {Kathiravan}}]{Har2015}
{Hariharan}, K., {Ramesh}, R., \& {Kathiravan}, C. 2015, Sol. Phys., 290, 2479,
  \dodoi{10.1007/s11207-015-0761-5}

\bibitem[{{Hariharan} {et~al.}(2014){Hariharan}, {Ramesh}, {Kishore},
  {Kathiravan}, \& {Gopalswamy}}]{Har2014}
{Hariharan}, K., {Ramesh}, R., {Kishore}, P., {Kathiravan}, C., \&
  {Gopalswamy}, N. 2014, APJ, 795, 14, \dodoi{10.1088/0004-637X/795/1/14}

\bibitem[{Huang {et~al.}(2018)Huang, Aa, Shen, \& Liu}]{Huang2018}
Huang, W., Aa, E., Shen, H., \& Liu, S. 2018, GPS Solutions, 22,
  \dodoi{10.1007/s10291-018-0780-4}

\bibitem[{{James} \& {Subramanian}(2018)}]{James2018}
{James}, T., \& {Subramanian}, P. 2018, MNRAS, 479, 1603,
  \dodoi{10.1093/mnras/sty1216}

\bibitem[{{James} {et~al.}(2017){James}, {Subramanian}, \&
  {Kontar}}]{James2017}
{James}, T., {Subramanian}, P., \& {Kontar}, E.~P. 2017, MNRAS, 471, 89,
  \dodoi{10.1093/mnras/stx1460}

\bibitem[{{Kishore} {et~al.}(2016){Kishore}, {Ramesh}, {Hariharan},
  {Kathiravan}, \& {Gopalswamy}}]{Kis2016}
{Kishore}, P., {Ramesh}, R., {Hariharan}, K., {Kathiravan}, C., \&
  {Gopalswamy}, N. 2016, APJ, 832, 59, \dodoi{10.3847/0004-637X/832/1/59}

\bibitem[{{Krypiak-Gregorczyk}(2019)}]{Krypiak2019}
{Krypiak-Gregorczyk}, A. 2019, Journal of Geodesy, 93, 931,
  \dodoi{10.1007/s00190-018-1216-1}

\bibitem[{{Kundu}(1965)}]{Kundu1965}
{Kundu}, M.~R. 1965, {Solar radio astronomy}

\bibitem[{{Liu} {et~al.}(2018){Liu}, {Chen}, {Cho}, {Feng}, {Vasanth}, {Koval},
  {Du}, {Wu}, \& {Li}}]{Liu2018}
{Liu}, H., {Chen}, Y., {Cho}, K., {et~al.} 2018, \solphys, 293, 58,
  \dodoi{10.1007/s11207-018-1280-y}

\bibitem[{{Lobzin} {et~al.}(2011){Lobzin}, {Cairns}, \&
  {Robinson}}]{Lobzin2011}
{Lobzin}, V., {Cairns}, I.~H., \& {Robinson}, P.~A. 2011, APJ, 736, L20,
  \dodoi{10.1088/2041-8205/736/1/L20}

\bibitem[{{Mahender} {et~al.}(2020){Mahender}, {Sasikumar Raja}, {Ramesh},
  {Pand iti}, {Monstein}, \& {Ganji}}]{Mah2020}
{Mahender}, A., {Sasikumar Raja}, K., {Ramesh}, R., {et~al.} 2020, arXiv
  e-prints, arXiv:2009.05755.
\newblock \doarXiv{2009.05755}

\bibitem[{{McLean} \& {Labrum}(1985)}]{McLean1985}
{McLean}, D.~J., \& {Labrum}, N.~R. 1985, {Solar radiophysics : studies of
  emission from the sun at metre wavelengths}

\bibitem[{{Melrose}(1980)}]{Melrose1980}
{Melrose}, D.~B. 1980, {Plasma astrohysics. Nonthermal processes in diffuse
  magnetized plasmas - Vol.1: The emission, absorption and transfer of waves in
  plasmas; Vol.2: Astrophysical applications}

\bibitem[{{Mercier}(1975)}]{Mercier1975}
{Mercier}, C. 1975, Sol. Phys., 45, 169, \dodoi{10.1007/BF00152229}

\bibitem[{{Mugundhan} {et~al.}(2017){Mugundhan}, {Hariharan}, \&
  {Ramesh}}]{Mugundhan2017}
{Mugundhan}, V., {Hariharan}, K., \& {Ramesh}, R. 2017, Sol. Phys., 292, 155,
  \dodoi{10.1007/s11207-017-1181-5}

\bibitem[{NCEI(2017)}]{NCEI2017}
NCEI. 2017

\bibitem[{{Nelson} \& {Melrose}(1985)}]{Nelson1985}
{Nelson}, G.~J., \& {Melrose}, D.~B. 1985, {Type II bursts.}, ed. D.~J.
  {McLean} \& N.~R. {Labrum}, 333--359

\bibitem[{{Nindos} {et~al.}(2011){Nindos}, {Alissandrakis}, {Hillaris}, \&
  {Preka-Papadema}}]{Nindos2011}
{Nindos}, A., {Alissandrakis}, C.~E., {Hillaris}, A., \& {Preka-Papadema}, P.
  2011, A\&A, 531, A31, \dodoi{10.1051/0004-6361/201116799}

\bibitem[{{Nindos} {et~al.}(2008){Nindos}, {Aurass}, {Klein}, \&
  {Trottet}}]{Nindos2008}
{Nindos}, A., {Aurass}, H., {Klein}, K.~L., \& {Trottet}, G. 2008, Sol. Phys.,
  253, 3, \dodoi{10.1007/s11207-008-9258-9}

\bibitem[{Patsourakos {et~al.}(2016)Patsourakos, Georgoulis, Vourlidas, Nindos,
  Sarris, Anagnostopoulos, Anastasiadis, Chintzoglou, Daglis, Gontikakis,
  Hatzigeorgiu, Iliopoulos, Katsavrias, Kouloumvakos, Moraitis,
  Nieves-Chinchilla, Pavlos, Sarafopoulos, Syntelis, Tsironis, Tziotziou,
  Vogiatzis, Balasis, Georgiou, Karakatsanis, Malandraki, Papadimitriou,
  Odstr{\v{c}}il, Pavlos, Podlachikova, Sandberg, Turner, Xenakis, Sarris,
  Tsinganos, \& Vlahos}]{Patsourakos2016}
Patsourakos, S., Georgoulis, M.~K., Vourlidas, A., {et~al.} 2016, AJ, 817, 14,
  \dodoi{10.3847/0004-637x/817/1/14}

\bibitem[{{Payne-Scott} {et~al.}(1947){Payne-Scott}, {Yabsley}, \&
  {Bolton}}]{Pay1947}
{Payne-Scott}, R., {Yabsley}, D.~E., \& {Bolton}, J.~G. 1947, Nature, 160, 256,
  \dodoi{10.1038/160256b0}

\bibitem[{{Pick} \& {Keller}(2004)}]{Pick2004}
{Pick}, D.~E., \& {Keller}, C.~U. 2004, {In: D. E. Gary and C. U. Keller
  (eds.): Solar and Space Weather Radiophysics. Dordrecht, pp. 17–45.}, Vol.
  314

\bibitem[{{Pick} \& {Ji}(1986)}]{Pick1986}
{Pick}, M., \& {Ji}, S.~C. 1986, Sol. Phys., 107, 159,
  \dodoi{10.1007/BF00155349}

\bibitem[{{Pohjolainen} {et~al.}(2007){Pohjolainen}, {van Driel-Gesztelyi},
  {Culhane}, {Manoharan}, \& {Elliott}}]{Poh2007}
{Pohjolainen}, S., {van Driel-Gesztelyi}, L., {Culhane}, J.~L., {Manoharan},
  P.~K., \& {Elliott}, H.~A. 2007, Sol. Phys., 244, 167,
  \dodoi{10.1007/s11207-007-9006-6}

\bibitem[{{Prieto} {et~al.}(2020){Prieto}, {Gordo}, {Rodr{\'\i}guez-Pacheco},
  {Mart{\'\i}nez}, {S{\'a}nchez}, {Russu}, {Monstein}, \&
  {Fern{\'a}ndez}}]{Prieto2020}
{Prieto}, M., {Gordo}, J.~B., {Rodr{\'\i}guez-Pacheco}, J., {et~al.} 2020, Sol.
  Phys., 295, 11, \dodoi{10.1007/s11207-019-1577-5}

\bibitem[{{Ramesh} {et~al.}(2010){Ramesh}, {Kathiravan}, {Barve}, {Beeharry},
  \& {Rajasekara}}]{Ramesh2010}
{Ramesh}, R., {Kathiravan}, C., {Barve}, I.~V., {Beeharry}, G.~K., \&
  {Rajasekara}, G.~N. 2010, APJ, 719, L41, \dodoi{10.1088/2041-8205/719/1/L41}

\bibitem[{{Ramesh} {et~al.}(2013){Ramesh}, {Sasikumar Raja}, {Kathiravan}, \&
  {Narayanan}}]{Ramesh2013}
{Ramesh}, R., {Sasikumar Raja}, K., {Kathiravan}, C., \& {Narayanan}, A.~S.
  2013, ApJ, 762, 89, \dodoi{10.1088/0004-637X/762/2/89}

\bibitem[{Redmon {et~al.}(2018)Redmon, Seaton, Steenburgh, He, \&
  Rodriguez}]{Redmon2018}
Redmon, R.~J., Seaton, D.~B., Steenburgh, R., He, J., \& Rodriguez, J.~V. 2018,
  Space Weather, 16, 1190, \dodoi{10.1029/2018SW001897}

\bibitem[{{Reid} \& {Ratcliffe}(2014)}]{Reid2014}
{Reid}, H.~A.~S., \& {Ratcliffe}, H. 2014, Research in Astronomy and
  Astrophysics, 14, 773, \dodoi{10.1088/1674-4527/14/7/003}

\bibitem[{Saint-Hilaire {et~al.}(2012)Saint-Hilaire, Vilmer, \&
  Kerdraon}]{Saint_Hilaire_2012}
Saint-Hilaire, P., Vilmer, N., \& Kerdraon, A. 2012, APJ, 762, 60,
  \dodoi{10.1088/0004-637x/762/1/60}

\bibitem[{{Sasikumar Raja} \& {Ramesh}(2013)}]{Sasikumar2013}
{Sasikumar Raja}, K., \& {Ramesh}, R. 2013, ApJ, 775, 38,
  \dodoi{10.1088/0004-637X/775/1/38}

\bibitem[{{Sasikumar Raja} {et~al.}(2014){Sasikumar Raja}, {Ramesh},
  {Hariharan}, {Kathiravan}, \& {Wang}}]{Sas2014}
{Sasikumar Raja}, K., {Ramesh}, R., {Hariharan}, K., {Kathiravan}, C., \&
  {Wang}, T.~J. 2014, APJ, 796, 56, \dodoi{10.1088/0004-637X/796/1/56}

\bibitem[{{Sasikumar Raja} {et~al.}(2018){Sasikumar Raja}, {Subramanian},
  {Ananthakrishnan}, \& {Monstein}}]{Sas2018}
{Sasikumar Raja}, K., {Subramanian}, P., {Ananthakrishnan}, S., \& {Monstein},
  C. 2018, arXiv e-prints, arXiv:1801.03547.
\newblock \doarXiv{1801.03547}

\bibitem[{Sato {et~al.}(2019)Sato, Jakowski, Berdermann, Jiricka, Heßelbarth,
  Banyś, \& Wilken}]{Sato2019}
Sato, H., Jakowski, N., Berdermann, J., {et~al.} 2019, Space Weather, 17, 816,
  \dodoi{10.1029/2019SW002198}

\bibitem[{{Selvakumaran} {et~al.}(2015){Selvakumaran}, {Maurya}, {Gokani},
  {Veenadhari}, {Kumar}, {Venkatesham}, {Phanikumar}, {Singh}, {Siingh}, \&
  {Singh}}]{Sel2015}
{Selvakumaran}, R., {Maurya}, A.~K., {Gokani}, S.~A., {et~al.} 2015, Journal of
  Atmospheric and Solar-Terrestrial Physics, 123, 102,
  \dodoi{10.1016/j.jastp.2014.12.009}

\bibitem[{{Sharma} {et~al.}(2018){Sharma}, {Oberoi}, \&
  {Arjunwadkar}}]{Sharma2018}
{Sharma}, R., {Oberoi}, D., \& {Arjunwadkar}, M. 2018, ApJ, 852, 69,
  \dodoi{10.3847/1538-4357/aa9d96}

\bibitem[{{Singh} {et~al.}(2019){Singh}, {Sasikumar Raja}, {Subramanian},
  {Ramesh}, \& {Monstein}}]{Dayal2019}
{Singh}, D., {Sasikumar Raja}, K., {Subramanian}, P., {Ramesh}, R., \&
  {Monstein}, C. 2019, Sol. Phys., 294, 112, \dodoi{10.1007/s11207-019-1500-0}

\bibitem[{{Stewart} {et~al.}(1982){Stewart}, {Dulk}, {Sheridan}, {House},
  {Wagner}, {Illing}, \& {Sawyer}}]{Ste1982}
{Stewart}, R.~T., {Dulk}, G.~A., {Sheridan}, K.~V., {et~al.} 1982, A\&A, 116,
  217

\bibitem[{{Tsurutani} {et~al.}(2014){Tsurutani}, {Echer}, {Shibata},
  {Verkhoglyadova}, {Mannucci}, {Gonzalez}, {Kozyra}, \&
  {P{\"a}tzold}}]{Tsurutani2014}
{Tsurutani}, B.~T., {Echer}, E., {Shibata}, K., {et~al.} 2014, JSWSC, 4, A02,
  \dodoi{10.1051/swsc/2013056}

\bibitem[{{Vasanth} {et~al.}(2016){Vasanth}, {Chen}, {Feng}, {Ma}, {Du},
  {Song}, {Kong}, \& {Wang}}]{Vas2016}
{Vasanth}, V., {Chen}, Y., {Feng}, S., {et~al.} 2016, \apjl, 830, L2,
  \dodoi{10.3847/2041-8205/830/1/L2}

\bibitem[{{Vasanth} {et~al.}(2019){Vasanth}, {Chen}, {Lv}, {Ning}, {Li},
  {Feng}, {Wu}, \& {Du}}]{Vas2019}
{Vasanth}, V., {Chen}, Y., {Lv}, M., {et~al.} 2019, \apj, 870, 30,
  \dodoi{10.3847/1538-4357/aaeffd}

\bibitem[{{Vasanth} {et~al.}(2014){Vasanth}, {Umapathy}, {Vr{\v{s}}nak},
  {{\v{Z}}ic}, \& {Prakash}}]{Vas2014}
{Vasanth}, V., {Umapathy}, S., {Vr{\v{s}}nak}, B., {{\v{Z}}ic}, T., \&
  {Prakash}, O. 2014, Sol. Phys., 289, 251, \dodoi{10.1007/s11207-013-0318-4}

\bibitem[{{Vourlidas} {et~al.}(2020){Vourlidas}, {Carley}, \&
  {Vilmer}}]{Vor2020}
{Vourlidas}, A., {Carley}, E.~P., \& {Vilmer}, N. 2020, Frontiers in Astronomy
  and Space Sciences, 7, 43, \dodoi{10.3389/fspas.2020.00043}

\bibitem[{{Vr{\v{s}}nak} {et~al.}(2001){Vr{\v{s}}nak}, {Aurass},
  {Magdaleni{\'c}}, \& {Gopalswamy}}]{Vrs2001}
{Vr{\v{s}}nak}, B., {Aurass}, H., {Magdaleni{\'c}}, J., \& {Gopalswamy}, N.
  2001, A\&A, 377, 321, \dodoi{10.1051/0004-6361:20011067}

\bibitem[{{Vr{\v{s}}nak} \& {Cliver}(2008)}]{Vrsnak2008}
{Vr{\v{s}}nak}, B., \& {Cliver}, E.~W. 2008, Sol. Phys., 253, 215,
  \dodoi{10.1007/s11207-008-9241-5}

\bibitem[{White(2007)}]{White2007}
White, S. 2007, Asian J. Phys., 16

\bibitem[{{Wild}(1950)}]{Wild1950}
{Wild}, J.~P. 1950, Australian Journal of Scientific Research A Physical
  Sciences, 3, 541, \dodoi{10.1071/PH500541}

\bibitem[{{Wild} {et~al.}(1963){Wild}, {Smerd}, \& {Weiss}}]{Wild1963}
{Wild}, J.~P., {Smerd}, S.~F., \& {Weiss}, A.~A. 1963, ARA\&A, 1, 291,
  \dodoi{10.1146/annurev.aa.01.090163.001451}

\bibitem[{{Zhang} {et~al.}(2018){Zhang}, {Wang}, \& {Ye}}]{Zhang2018}
{Zhang}, P.~J., {Wang}, C.~B., \& {Ye}, L. 2018, A\&A, 618, A165,
  \dodoi{10.1051/0004-6361/201833260}

\bibitem[{{Zheleznyakov}(1970)}]{Zheleznyakov1970}
{Zheleznyakov}, V.~V. 1970, {Radio emission of the sun and planets}

\bibitem[{{Zucca} {et~al.}(2012){Zucca}, {Carley}, {McCauley}, {Gallagher},
  {Monstein}, \& {McAteer}}]{Zuc2012}
{Zucca}, P., {Carley}, E.~P., {McCauley}, J., {et~al.} 2012, Sol. Phys., 280,
  591, \dodoi{10.1007/s11207-012-9992-x}

\end{thebibliography}

\end{document}